\documentclass[useAMS,usenatbib]{mn2e}

\usepackage{times}            
\usepackage{graphicx,epsfig}
\usepackage{multicol}
\usepackage{floatflt}

\def \gsim { \lower .75ex \hbox{$\sim$} \llap{\raise .27ex \hbox{$>$}} }
\def \lsim { \lower .75ex \hbox{$\sim$} \llap{\raise .27ex \hbox{$<$}} }
\def \deg {^{\circ}}

\def \WMAP5 {{\it WMAP5}}

\def\be{\begin{equation}}
\def\ee{\end{equation}}
\title[WMAP beam sensitivity]{Beam profile sensitivity of the WMAP  CMB power spectrum}
\author[U. Sawangwit \& T. Shanks]{U. Sawangwit\thanks{E-mail: utane.sawangwit@durham.ac.uk, tom.shanks@durham.ac.uk} and  T. Shanks\footnotemark[1]\\
Department of Physics, Durham University, South Road, Durham, DH1 3LE, UK.}
\begin{document}
\date{Accepted 2010 June 2.  Received 2010 May 28; in original form 2009 December 2}
\pagerange{\pageref{firstpage}--\pageref{lastpage}} \pubyear{2010}
\maketitle
\label{firstpage}
\begin{abstract}
Using the published WMAP 5-year data, we  first show how sensitive the WMAP
power spectra are to the form of the WMAP beam. It is well known that the beam
profile derived from observations of Jupiter is non-Gaussian and indeed extends,
in the W band for example, well beyond its $12.'6$ FWHM core out to more than 1
degree in radius. This means that even though the core width corresponds to wavenumber
$l\approx1800$, the form of the beam still significantly affects the WMAP
results even at $l\approx200$ which is the scale of the first acoustic peak. The
difference between the beam convolved $C_l$ and the final $C_l$ is $\approx70$\% at
the scale of the first peak, rising to $\approx400$\% at the scale of the second.

New estimates of the Q, V and W-band beam profiles are then presented, based on
a stacking analysis of the WMAP5 radio source catalogue and temperature maps. 
The radio sources show a significantly ($3-4\sigma$) broader beam profile 
on scales of $10'-30'$ than that found by the WMAP team whose beam analysis is based 
on measurements of Jupiter. Beyond these scales the beam profiles from the radio 
sources are too noisy to give useful information. Furthermore, we find tentative 
evidence for a non-linear relation between WMAP and ATCA/IRAM 95 GHz source fluxes. 
We discuss whether the wide beam profiles could be caused either by radio source 
extension or clustering and find that neither explanation is likely. 
We also argue against the possibility that Eddington bias is affecting our results. 
The reasons for the difference between the radio source and the Jupiter beam profiles are 
therefore still unclear. If the radio source profiles were then used to define the WMAP beam, 
there could be a significant change in the amplitude and position of even the first acoustic 
peak. It is therefore important to identify the reasons for the differences between these two 
beam profile estimates.
\vspace{-1.5mm} 
\end{abstract}
\begin{keywords}
cosmic microwave background -- cosmology: observations-early universe--space vehicles: instruments
\vspace{-10mm}
\end{keywords}
\section{Introduction}
\label{sec:intro}
The WMAP satellite has produced some of the best support for the standard
$\Lambda$CDM cosmological model. By measuring the first two acoustic peaks it has shown that
the Universe is spatially flat with $\Omega_\Lambda=0.74$ and
$H_0=72$ kms$^{-1}$Mpc$^{-1}$ \citep{Hinshaw09}. The precision of the fit is 
impressive and rules out many competing simple models such as the low $H_0$, 
$\Omega_{\rm baryon}=1$ model of \citet{Shanks85,Shanks05,Shanks07}.

Of course, statistically precise measurements can also contain systematic errors
which have to be guarded against. Such systematics include Galactic foregrounds 
which at the least cause mode coupling due to the incomplete sky \citep[e.g.][]{Hinshaw03,Chon04}. 
There are also potentially more subtle systematics that arise from
cosmological foregrounds. For example, \cite{Myers04} and \cite{Bielby07} have 
detected the SZ effect in the WMAP data by cross-correlating the CMB
with rich cluster positions. \cite{Shanks07} has also discussed the effect of
foreground lensing, prompted by QSO lensing results \citep{Myers03,Myers05,Mountrichas07}. 
But SZ is unlikely to make a strong contribution to the first acoustic peak 
\citep{Huffenberger04}. Also lensing requires a high anti-bias between galaxies and 
the mass distribution to have a significant effect at the first peak which needs to be 
reconciled with measures of bias from galaxy clustering dynamics 
\citep[e.g.][]{Ratcliffe98,Hawkins03}.

However, there are also many potential systematics involved with the WMAP
instrument, although the WMAP team have taken care that the effects of such
systematics are minimised. One major potential systematic concerns the question
of the WMAP radio telescope beam profile. We shall see that even at the wavenumber
$l\approx220$ of the first acoustic peak, the CMB power spectrum has significant
dependence on the beam profile even in the highest resolution W band. Here
the W-band resolution quoted by the WMAP team is $12.'6$ FWHM which is roughly
equivalent to $l\approx1800$. It is also noted that the beam is not Gaussian.
Now the WMAP team have extensive papers devoted to the important question of
measuring the beam \citep{Page03,Jarosik07,Hill09}. The standard method is
to use their observations of bright sources such as the planet Jupiter to
measure the beam profile.

Here, after describing the WMAP5 data in \S \ref{sec:data}, we re-derive in
\S \ref{sec:raw_cl} the raw CMB power spectrum from the WMAP maps to show directly
the effect of the beam. Then in \S \ref{sec:testing} we use radio sources to make new
estimates of the WMAP beam and discuss other tests of the beam profile. In
\S \ref{sec:debeam} we then make fits to the radio source beam profiles and use
these to de-beam the WMAP5 data and explore the range of power spectra
that results. Possible reasons for the beam profile discrepancy and our 
conclusions are then presented in \S \ref{sec:discussion}.
 \vspace{-5mm}
\section{Data}   
\label{sec:data}
\subsection {WMAP5 maps and point source catalogue} 
Here we use the five-year WMAP datasets which are available from the LAMBDA CMB
website. The maps from  the individual detectors in 5 frequency bands, K, Ka, Q,
V and W are supplied. The FWHM of the 94 GHz W beam is $12.'6$
compared to $19.'8$ at V (61 GHz), $29.'4$ at Q (41GHz), $37.'2$ at Ka (33GHz)
and  $49.'2$ at K (23GHz). There are 10 differencing assemblies (DAs), namely 
K1, Ka1, Q1, Q2, V1, V2, W1, W2, W3 and W4. The different DA maps can be
cross-correlated to obtain power spectra free of uncorrelated detector noise bias 
\citep{Hinshaw03}. The Jupiter beam profiles for each DA and the corresponding 
transfer functions are also given. The maps are in HEALPix \citep{HEALPixref} 
format with $N_{\rm side}=512$ and $N_{\rm side}=1024$. These give equal area pixels of 
dimension $\approx7'$ and $\approx3'$, respectively.

We use the radio sources drawn from the WMAP5 point source catalogue \citep{Wright09}. 
These sources have to be detected to $>5\sigma$ in at least one WMAP band and 
their flux density is reported if they are detected at $>2\sigma$ in
any of the other four WMAP bands. This gives a list of 390 sources to a
limit of $\approx$0.5 Jy in each band. The source positions are accurate to $\sim 4'$ \citep{Wright09}. 
365 out of 390 sources are pre-detected at 4.85 GHz in the Greenbank (GB6) northern sky survey 
\citep{GB6ref} and the Parkes-MIT-NRAO (PMN) surveys \citep{PMNref}. Here, we only use WMAP5 sources with 
4.85 GHz counterparts and exclude sources (12 out of 365) that were found to be resolved 
at $\approx4.'6$ FWHM resolution of GB6 and PMN. Table 1 shows the number of these sources 
in each band, also split into those brighter or fainter than 1.1 Jy.

\begin{table}
\centering
\caption{Summary of the WMAP sources listed as point sources in the Greenbank and PMN 5GHz 
catalogues.}
	\vspace{-2mm}
	\begin{tabular}{lccc}
        \hline
        \hline
Band	    & $\ge$1.1Jy  &$<1.1$Jy&	Total ($>2\sigma$) \\
           \hline
Q	&182& 165 &347   \\
V 	&164& 153 &317   \\
W       &97 &  84 &181   \\
	\hline
	\hline
		\end{tabular}
\label{tab:summary}
\vspace{-3mm}
\end{table}

From the optical identifications of \cite{Trushkin03} of the 208 WMAP 1st year
sources the survey contains 77 per\,cent QSOs or BL Lac with the remainder being radio
galaxies/AGN. This is as expected given the dominance of flat-spectrum compact
sources at the high WMAP frequencies. 
\vspace{-4mm}
\subsection{Ground-based 90-95GHz Radio Sources}
\label{sec:atcairam}
We shall compare WMAP W-band fluxes with ground-based radio source fluxes from
ATCA \citep{Sadler08} and IRAM \citep{Steppe88}. The ATCA survey was made
at 95 GHZ and the IRAM survey at 90GHz. The ATCA survey was based on sources
selected at 20GHz. Of the 130 sources observed, 17 were detected at more than
$2\sigma$ by WMAP5. The IRAM survey observed 294 sources at 90
and 230 GHz, targetting sources which are brighter than 1 Jy at 5 GHz. Here 66
sources were detected at more than $2\sigma$ by WMAP5. At these high frequencies
the sources are mainly QSOs, BL Lacs or blazars. Many of the sources in the ATCA
and IRAM surveys are variable and so where this is an issue we shall use the
average source fluxes in our comparison with the WMAP fluxes.

\begin{figure}
\hspace{-7mm}
\includegraphics[scale=0.5]{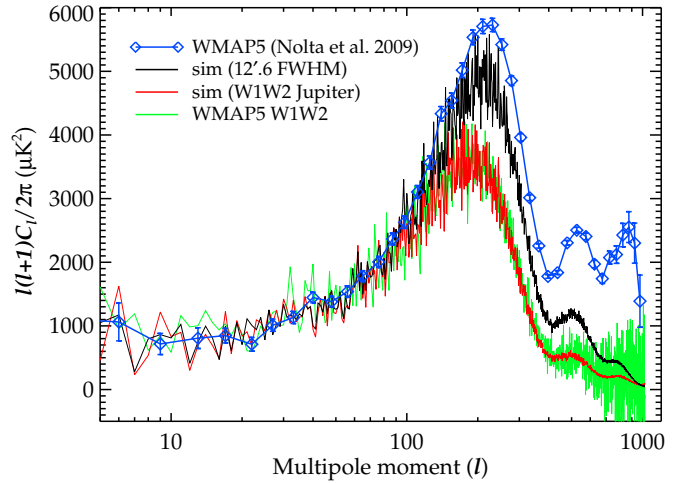}
\vspace{-7mm}
\caption{The beam convolved $C_l$ measured from the W1 and W2 maps (green line)
compared to the standard de-beamed WMAP result as presented by Nolta et al. (blue line). 
Also shown are $C_l$ measurements of a full-sky CMB simulation (WMAP5 best-fit $\Lambda$CDM) 
smoothed with a $12.'6$ FWHM Gaussian beam (black line), and a
similar simulation now smoothed with the W1 and W2 Jupiter beams (red line). 
Although the beam convolved $C_l$ (green) and the Jupiter beam-smoothed
standard simulation $C_l$ (red) agree, the difference between these
and the Nolta et al result (blue) shows the large effect of debeaming
even at the scale of the first acoustic peak.}
	\label{fig:beameffect}
	\vspace{-2mm}
\end{figure}
\vspace{-5mm}
\section{Deriving the beam convolved $C_{\lowercase{l}}$} 
\label{sec:raw_cl}
\begin{figure*}
 	\hspace{-3mm}
\includegraphics[scale=0.51]{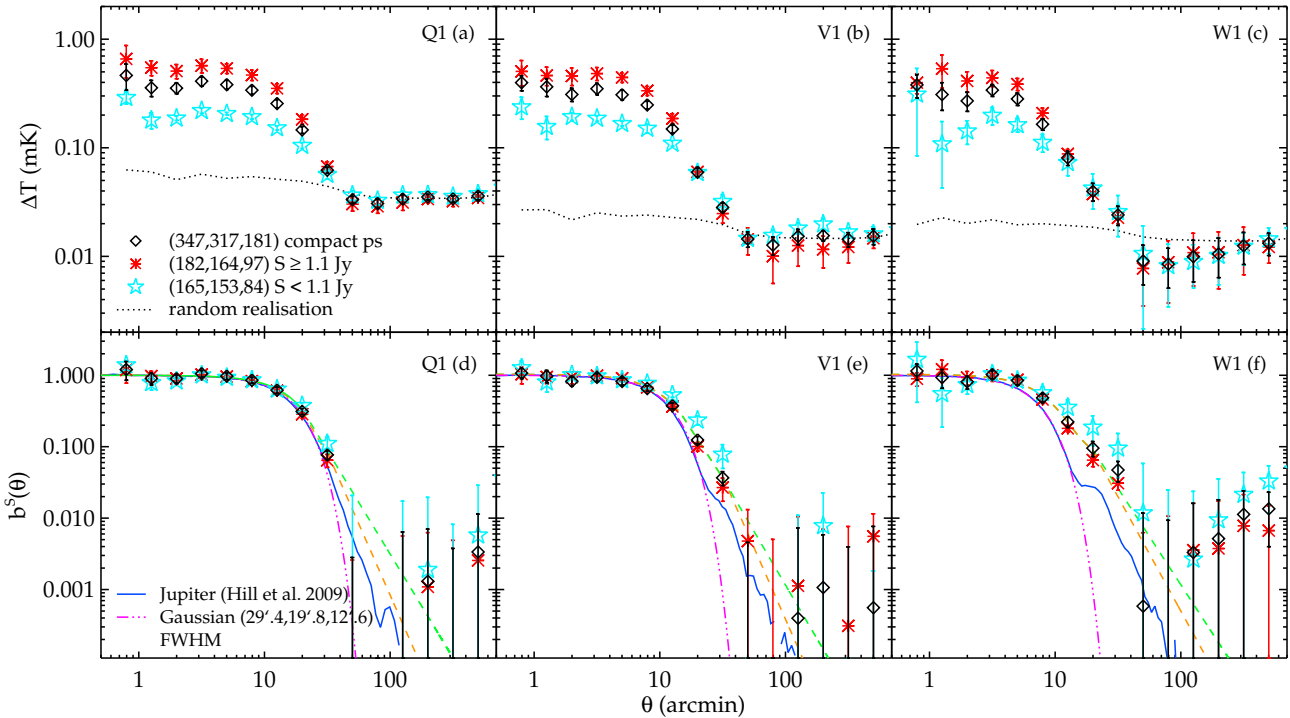}
\vspace{-5mm}
\caption{(a), (b) and (c); The raw radio source beam profiles for the WMAP sources compared
to the beam profiles from Jupiter (solid lines) and random realisations (dotted lines) 
for Q1, V1 and W1. (d), (e) and (f); The zero-offset subtracted (see text) and normalised beam 
profiles for Q1, V1 and W1. The radio source profiles for all 
compact WMAP sources are shown as diamonds. Profiles derived from WMAP sources with flux 
brighter (fainter) than $1.1$ Jy are shown as asterisks (stars). 
A Gaussian is shown as a dot-dashed line and empirical fits to the radio source profiles 
are shown as orange and green dashed lines.}
	\label{fig:radiobeam}
	\vspace{-5.0mm}
\end{figure*}

We now analyse the WMAP W-band data to make an initial estimate of the power
spectrum from the W band. To reduce the effect of correlated detector noise
which would result from an auto-correlation of an individual CMB map, we make a
cross-correlation of the maps from independent detectors. We derive the result
by using the {\it PolSpice} code \citep{Szapudi01} to cross-correlate the W1 and W2 
maps. The default WMAP5 temperature power spectrum mask \citep[KQ85,][]{Nolta09} is 
used and the cut-sky corrected angular power spectrum, $C'_l$, is obtained directly 
from the {\it PolSpice} code \citep[see][]{Chon04}. Hereafter, we shall call an 
angular power spectrum after a cut-sky and pixel transfer function (see Eq. \ref{eq:cl}) 
correction a `beam convolved $C_l$'. 
In Fig. \ref{fig:beameffect}, we immediately see that the beam convolved $C_l$ (green line) 
is not only drastically smoothed at the position of the second and third peaks but there is
also a significant effect at the position of the first peak at $l\approx220$ in
that the amplitude of the standard $\Lambda$CDM result (blue line) is $\approx70$\% higher. 
The reason for this is seen in Fig. \ref{fig:radiobeam}f where 
the beam profile from the Jupiter observations using the W1 detector are shown. 
It can be seen that the beam is not Gaussian and has a $\theta^{-3}$ power-law tail out to
$>1\deg$.

\citet{Page03} give the relation between the beam transfer function, $b_l$,
and the radial beam profile, $b^S(\theta)$, as,

\begin{equation}
b_l=2\pi\int{b^S(\theta)P_l(\cos\theta)d \cos\theta}/\Omega_B
\label{eq:bl}
\end{equation}

\noindent The de-beamed cross-power spectrum measured from DA $i$ and $j$ is then given from 
the measured $C'_l$ by 
\begin{equation}
 C_l=C'_l/b_l^i b_l^j p_l^2,
\label{eq:cl}
\end{equation}
where $p_l$ is the pixel transfer function supplied with the HEALPix package. For $N_{\rm side}=512$, 
the pixelisation lowers the measured $C'_l$ by $\approx1$ and 10 per\,cent at $l\approx200$ and 
500, respectively.     

If we use Eq. \ref{eq:cl} with the Jupiter beam transfer function from the WMAP team, 
we find that we get back to the usual $\Lambda$CDM model (green and orange 
lines in Fig. \ref{fig:debeam}). 
The black line shows the $C_l$ measured from a full-sky CMB simulation (WMAP5 best-fit $\Lambda$CDM model) 
after smoothing by a Gaussian beam, using {\it synfast} code \citep{HEALPixref}. 
The red line shows the effect of the similar simulation now smoothed with W1W2 Jupiter beam profiles. 
The latter shows excellent agreement with our beam convolved $C_l$ measured from the W1 and W2 maps. 
The W1W2 $C_l$ is noisier than the simulation due to radiometer noise in the data. 
The effect of the Jupiter beam compared to the Gaussian beam is thus very significant 
in decreasing the height of the first peak. We also see that when the Jupiter beam is used, 
the $\Lambda$CDM model does give an accurate fit to the beam-convolved $C_l$. Thus when we use the same 
parameters as the WMAP team, we reproduce the WMAP result.

\vspace{-5mm}
\section{Testing the WMAP beam profile}
\label{sec:testing}
\subsection{Beam profiles via point sources} 
\label{sec:method}
We then estimated a beam from the radio sources by making a stacking
analysis of WMAP5 temperature maps around radio source positions. 
The extended foreground emission regions are excluded from the temperature maps using the 
`point source catalog mask' \citep{Wright09}.  
We calculated the average $\Delta T$ (per 49$\,$arcmin$^2$ pixel) in annuli as 
a function of angular distance, $\theta$, between radio source position and the
pixel centre. In the first instance we show the raw cross-correlation function
for the Q, V and W bands in Fig. \ref{fig:radiobeam}, split into bright ($\ge1.1$ Jy) and faint
($<1.1$ Jy) WMAP5 source sub-samples. The errors on the radio source profiles 
are {\it jackknife} errors. These are estimated using six equal-area sub-fields, given 
by $\sigma_{\rm JK}^2(\theta)=(N_J-1) \langle [\Delta T(\theta)-\overline{\Delta T}(\theta)]^2 \rangle $, 
where $N_{J}=6$ and $\Delta T$ is the stacked temperature measured from five out of 
the six sub-fields. 

We see that the 
fainter source profiles appear to agree with the brighter source profiles 
at scales of $\theta\approx30'$ but have significantly lower peak values. 
This is most clearly shown in the un-renormalised profiles shown for the  
bright and faint Q,V, W band sources in Fig. \ref{fig:radiobeam}a, b and c. 
Although noise may be an issue for the faintest sources, this suggests that 
there may possibly be some form of non-linearity in the WMAP beam. We also 
note that the profiles from both bright and faint sources show a positive 
offset at the 0.01-0.02 mK level. The offset shows an increasing trend 
from $1\deg-5\deg$. The main uncertainty in estimating WMAP beam profiles 
from these data is in subtracting this offset at scales $>1\deg$.

Since WMAP has significant sidelobes stretching to $\approx90\deg$ \citep{Barnes03}, 
there was a possibility that the offsets are also part of the beam.
However, when we distributed points at random in the masked region and used these as
our centres for our stacking analysis we also found a similar offset
(dotted lines in Fig. \ref{fig:radiobeam}a, b and c). This makes it look like the 
offset is not associated with the existence of sources and hence not associated with 
the WMAP beam. Our Monte Carlo simulations (see below) shows that these offsets are 
caused by the CMB fluctuations. We therefore employed a `photometry' approach for 
the stacking analysis where we have subtracted the WMAP flux in an annulus at 
$1\deg<\theta<2\deg$ for the W band and proportionately bigger annuli in the V and Q bands. 
Using sky annuli close to the sources will clearly improve background subtraction in the
presence of local background fluctuations.

The resulting WMAP radio source beam profiles for Q, V and W bands are shown 
in Fig. \ref{fig:radiobeam}d, e and f. The profiles have been renormalised 
($\approx10$\% statistical uncertainties in the normalising factors) to fit the peak 
in the WMAP Jupiter beam profile at $\theta<4'$ and this profile is also shown. 
For each band we also 
compare the profiles to a Gaussian beam with the FWHM as indicated in the plot. 
We see that on average the profiles from the radio sources are broader than the
Jupiter profile in the W, V and Q bands. In the lower frequency, lower
resolution K and Ka bands the radio source profiles fit the Jupiter beam better,
indeed almost perfectly (not shown here). Clearly, given the size of the errors 
there is little information from the radio sources on the beam profile at $\theta>30'$. 
Fig. \ref{fig:radiobeam} again shows the WMAP radio sources divided into faint 
and bright sources, split at 1.1 Jy. In the W and V bands particularly we again 
note that the fainter sources appear to be wider than the brighter sources. We also find similar results 
for W2, W3, W4, V2 and Q2 but choose not to include them here for clarity. 
These deviations from the WMAP Jupiter beam are puzzling and we now check to 
see if they could be caused by systematics.

At the referee's request, we made 100 Monte Carlo simulations following \cite{Wright09} to check our results. 
In summary, simulated maps are constructed to include point sources sampled from a power law $N(>S)$ 
distribution with spectral characteristics as seen in the data. The temperature map for each band is then 
smoothed with the WMAP Jupiter beam profile before being added to a simulated CMB map including radiometer noise. 
We then applied the five-band detection method following procedures described by \cite{Wright09}. 
Applying our stacking analysis described above, we found that even profiles as narrow as the W-band 
Jupiter profile can be accurately retrieved. The flux 
dependence of measured profiles were small with only a hint of possible Eddington bias in the faintest bin. 
The pixelisation effect is also too small to 
explain the wider profiles seen here. Furthermore, the estimated uncertainties using these simulations 
are consistent with the Jackknife error estimates. Further details are given by \cite{SawangwitThesis} who, 
further, finds similarly wide profiles for NVSS radio sources flux limited at 1.4 GHz, 
where any Eddington bias would be negligible.

We have checked the likely contribution of radio source clustering to the beam
profiles, using the clustering analysis of the NVSS radio survey by \cite{Overzier03}. 
At $S\ge200$ mJy where the sky density of NVSS sources is $n\approx0.6$ deg$^{-2}$,

\begin{equation}
w(\theta)=3\times10^{-5}\theta^{-3.4}+6.6\times10^{-3}\theta^{-0.8}.
\label{eq:wtheta}
\end{equation}

\noindent This is a 2-power-law form which changes slope at
$\theta\approx0.1\deg$. At smaller scales, double-lobed radio sources split into 
two components dominate while at larger scales source-source clustering dominates.
We first calculate the excess number
of sources in an annulus of area $\Delta A$ at radius $\theta$ from an average 
source, $N_{ex}(\theta)=w(\theta)n\Delta A$. The excess flux/temperature per
unit area in the profile in the annulus is then given by $\Delta
T_{ex}=N_{ex}\times \bar{f} /\Delta A$ where $\bar{f}$ is the average source flux. For a
Gaussian point source of central intensity/temperature per unit area, $T_0$, and width, $\sigma$, 
the flux is $2\pi\sigma^2 T_0$. Therefore in this case, 
$\Delta T_{ex}(\theta)=w(\theta) 2\pi n\sigma^2 T_0$.
We find $\Delta T_{ex}\approx3\times10^{-4}T_0$
which is a negligible contribution in explaining the excess in Fig. \ref{fig:radiobeam}f at this
scale, if $T_0$ is taken to be the central profile value. Taking the parameters for 100 mJy from 
\cite{Overzier03} makes the effect
even smaller. These results are also likely to be upper limits for the WMAP
sources which only have a density of $n\approx0.01$ deg$^{-2}$ and an average 95 GHz
flux of 500 mJy. We conclude that radio source clustering is not likely to be an
issue for our radio source beam profiles.

\begin{figure}
\hspace{-0.5cm}
	\centering
	\hspace{-3mm}
\includegraphics[scale=0.47]{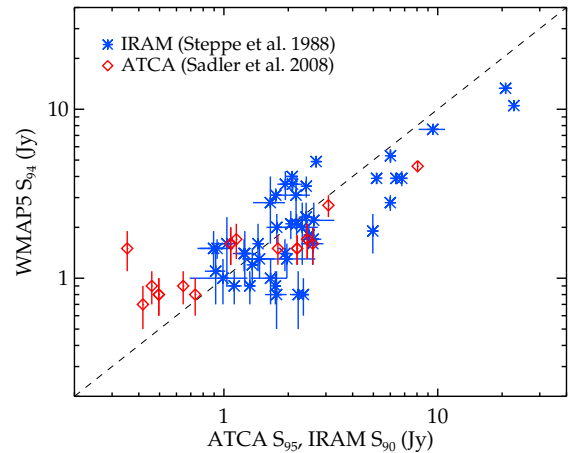}
\vspace{-4mm}
\caption{The comparison of the ATCA (diamonds) and IRAM (asterisks) source 
flux densities with the WMAP \textit{W}-band data.}
	\label{fig:flux}
	\vspace{-3mm}
\end{figure}

We conclude that in the W and V bands and probably the Q band, the average radio 
source profile is wider than the Jupiter beam and the fainter sources may show a 
wider profile than the brighter sources. 
For W1 and $S > 1.1$ Jy sources, the beam profile measured here rejects that of Jupiter 
with 4.0, 3.0 and 3.5$\sigma$ significance for $\theta=12.'6,20'$ and $31.'6$. These become 4.4, 
3.2 and 2.8$\sigma$ when Monte Carlo errors are used instead. Note that the pixelisation has been 
taken into account when estimating these significances.

\vspace{-5mm}
\subsection{Comparison with ground-based fluxes}
\label{sec:comparison}
We now make a check of the WMAP5 W band fluxes as presented by
\cite{Wright09} in their Table 1. We checked these against the ATCA and IRAM
source flux densities. The comparison in Fig. \ref{fig:flux} shows that for both surveys, the
brighter sources with fluxes $>3$ Jy are about a factor of 1.5 fainter in the
WMAP source list than in the ATCA or IRAM lists. The agreement between the ATCA
and IRAM fluxes appears better than for WMAP, if we use WMAP as an intermediary
between these two surveys. If the scale error is due to WMAP, then this might
suggest that there is a non-linearity in the WMAP flux scale. It could 
mean that a narrower WMAP beam at brighter fluxes is missing a significant
amount of flux in the tail of the beam profile.

\vspace{-5mm}
\section{Impact on the de-beamed $C_{\lowercase{l}}$} 
\label{sec:debeam}
Finally, we use the information from our radio source beam profiles to judge
what the effect might be on the de-beamed WMAP $C_l$. Unfortunately we will have
to extrapolate our radio source fits in the regime beyond $\sim1\deg$ out to $5\deg$
because of the large errors on the radio source beam profile in this range. These 
results can therefore only be used to indicate the sensitivity of the $C_l$ to the beam profile 
and should not be regarded as alternative $C_l$ estimates.
We first make an extrapolation where we fit the small-scale (total sample) beam profile points and
then extrapolate continuing with the power-law as shown by the green dashed lines in Fig. 
\ref{fig:radiobeam}d, e and f. 
We also made a more conservative extrapolation where we again fit the small-scale
data but then extrapolate continuing parallel to the Jupiter beam profiles at large
scales (orange dashed lines).

\begin{figure}

	\hspace{-7mm}
\includegraphics[scale=0.5]{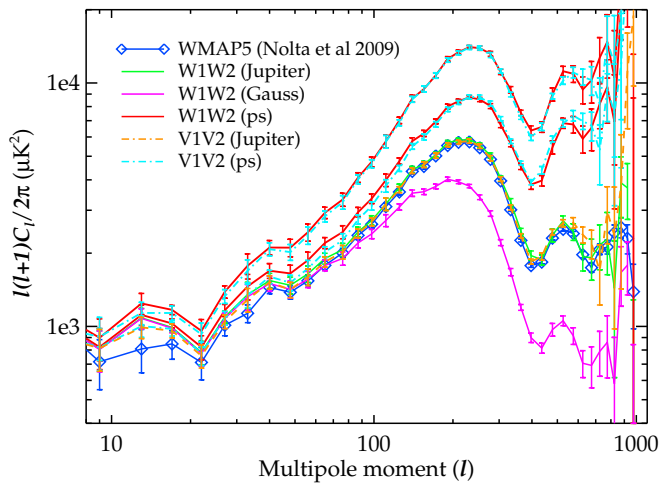}
\vspace{-7mm}
\caption{The de-beamed CMB power spectra. The blue diamonds show the WMAP team's result. 
The cross-$C_l$ for the W1 and W2 (V1 and V2) maps and de-beaming with the 
Jupiter beam profile is shown as the green (orange) line. The same result
but now de-beamed using the $12.'6$ FWHM Gaussian is shown as the magenta line, 
significantly different from the result using the Jupiter beam profile to de-beam. 
The same results but now using the profiles from the compact radio sources
are shown as the red and cyan lines for W and V band respectively. The differences in 
the two sets of de-beamed W and V spectra are due to the difference in extrapolations of 
the radio source beam profiles beyond $\theta=30'$ (see Fig. \ref{fig:radiobeam}e and 
\ref{fig:radiobeam}f.)}
	\label{fig:debeam}
	\vspace{-3mm}
\end{figure}

The range of the radio source de-beamed $C_l$ is shown by the two red lines and two 
cyan lines (for W and V bands) in Fig. \ref{fig:debeam}. The most conservative profile model is 
$\approx50$\% higher than the Jupiter de-beamed $C_l$ (green and orange lines) at the 
scale of the first peak. But the most extreme model is now a factor of 2-3 higher even 
at $l=220$ than the standard model power spectrum. We note that it has been possible to
derive consistent $C_l$'s between the V, W (and Q) bands, although we accept that this is
due to the freedom we have in extrapolating our radio source beam profiles beyond 
$\theta\approx30'$. It seems that if the radio sources are indicating a wider
beam profile, then the systematic uncertainty in the beam at the largest scales
will dominate the error budget of the $C_l$ even at the scale of the first
acoustic peak. These larger errors would then allow a wider range of 
cosmological models to be fitted, including models where the first peak lies at $l$ as high as 330 
\citep{Shanks10}.

\vspace{-5mm}
\section{Discussion and conclusions}
\label{sec:discussion}
Clearly it is important to understand why the radio source profiles are so wide
in the Q, V and W bands. If there is a correlation between beam width and source
flux then it will be wrong to use Jupiter to debeam the CMB power
spectra because in the W band, for example, the $\approx$1 Jy radio sources are
much closer to the $\approx0.5$ Jy rms flux of the CMB fluctuations than the $1200$ Jy
flux of Jupiter. 

The non-linearity shown by the WMAP source fluxes compared to
independently measured ATCA/IRAM fluxes is supporting evidence of
non-linearity in the WMAP data calibration. It is possible that
somehow the variability of the radio sources at W have caused problems that
would not apply to the CMB. In passing, we note that the smaller than expected SZ 
decrements from WMAP observations of rich clusters \citep{Myers04,Bielby07} may 
also be explained by a wider than expected WMAP beam at W. If so, this would argue 
that the non-linearity affects variable and non-variable WMAP sources alike. 

In considering possible 
causes of WMAP non-linearity, we first note that detector saturation is 
unlikely to be the problem since this would lead to
the brighter sources having a wider profile than the fainter sources, which is
opposite to what is observed. However, Jupiter, being a moving source, has to be
dealt with in a different way to the radio sources and the CMB fluctuations in
the maps. This means that if there was a problem in the WMAP analysis, it would be
necessary to check any filtering that is done to the maps. Otherwise, we do not 
understand the reason for the difference between the Jupiter and radio source 
beam profiles. 

We conclude that; 

\vspace{-2mm}
\begin{itemize}
\item The WMAP power spectrum is heavily dependent on the
beam profile. Indeed even the first acoustic peak at $l\approx220$ is very
dependent on the form of the profile at $1\deg-2\deg$ where the profile is only
$\approx0.1$ per\,cent of its peak value.
\item The radio point sources detected by WMAP in the Q, V and W bands
generally show a broader beam profile than the Jupiter beam used by the WMAP
team. For example, using bright point sources, our W1 beam profile rejects the Jupiter 
beam with $\ga$ 99.5\% confidence.    
\item There may be evidence for a flux dependent effect within the WMAP data
in that fainter radio sources appear to have systematically broader profiles than
brighter sources, although the faint data are noisy.
\item Non-linearity in the WMAP flux scale may also be indicated by
comparisons of WMAP radio source fluxes with ATCA and IRAM fluxes which show
50 per\,cent reduced flux from WMAP.
\item Further arguments against possible systematics such as Eddington bias affecting 
our results come from simulation checks and NVSS source samples selected at frequency 
where CMB fluctuations are subdominant (see Sawangwit 2010).  
\item The systematic 
errors on the WMAP $C_l$ due to the beam may be much larger than previously 
expected and in turn, this means that the systematic error on the best fit 
cosmological model may also be larger. It will be interesting to see if a
revised estimate of the WMAP beam profile then allows a simpler cosmological model 
to be fitted than $\Lambda$CDM.
\end{itemize}

\vspace{-0.5mm}
\noindent {\bf Acknowledgements:} We acknowledge the use of NASA WMAP data. 
We thank G. Hinshaw and E. Wright for useful comments. We thank an anonymous 
referee for very useful suggestions and comments. US acknowledges 
financial support from the Royal Thai Government.

\vspace{-7mm}

\bibliographystyle{mn2e}
\bibliography{beam}


\label{lastpage}
\end{document}